# A Review on understanding Brain, and Memory Retention and Recall Processes using EEG and fMRI techniques

Qazi Emad-ul-Haq[1], Muhammad Hussain[1], Hatim Aboalsamh[1], saeed bamatraf[1], Aamir Saeed Malik[2], Hafeez Ullah Amin[2]

[1]Department of Computer Science, College of Computer and Information Sciences, King Saud University, Riyadh, Saudi Arabia

[2]Centre for Intelligent Signal and Imaging Research (CISIR), Department of Electrical and Electronic Engineering, Universiti Teknologi PETRONAS, Seri Iskandar, Malaysia.

**Abstract:** Human memory - the learning of new information involves changes at the synaptic level between neurons dedicated for storage of information. Generally, memory is classified as Long-Term Memory and Short-Term Memory. The various types of the memory and their disorder are widely studied using neuroimaging techniques like Electroencephalography (EEG) and functional Magnetic Resonance Imaging (fMRI). Brain is effectively occupied with the capabilities of learning, retention and recall. The brain regions (pre-frontal cortex, associated hippocampus cortices and their interactions with other lobes) involved in memory recall tasks focuses on understanding the memory retention and recall processes. However, due to highly complicated and dynamic mechanisms of brain, the specific regions where information may reside are not completely explored. In this research paper, recent memory literature using EEG and fMRI studies is reviewed to understand the memory retention and recall processes as well as the various brain regions associated with these processes. A number of stimuli which are reported in previous studies are evaluated and discussed. Furthermore, the challenges which are being faced by researchers in EEG and fMRI methodologies are also presented. Recommendations for the future research related to memory retention and recall are also discussed at the end.

## 1. INTRODUCTION

Learning, memory retention and recall are the important cognitive processes which are deeply interlinked. But yet, there are prominent differences between them in the areas of neurosciences and cognitive neuropsychology. All our thoughts and actions depend on memory that remains dynamic constantly, getting fresh data from human senses, overhauling presented information, recovering the stored abilities and experiences, and anticipating upcoming actions that have not happened so far. Basically, learning is related with variations in behaviors which occurs due to experience. On the contrary, memory is related to the capability of retaining and recalling the learned information, i.e., memory processes consist of encoding, retention and recall. The encoding process in the human brain permits change over the apparent data of concern in a structure, which may be stored. That is the initial phase of making new memory in the human brain. The second memory process, i.e., retention is used to keep the encoded data. Recall is the third memory process which is used to retrieve the held data from the brain. Generally, there are two types of memories in human brain, i.e., Short Term Memory (STM) and Long Term Memory (LTM), which retains and recalls data in different ways, and a number of brain parts are actively involved simultaneously in this process. STM holds data temporarily to retain seven to nine items only in memory of an individual [1]. It has a tendency to be deteriorated as an individual gets older. On the contrary, LTM retains unlimited quantity of data for a longer period of time. That data may include, for example, home occasions, spatial and temporal relations between those occasions, their importance, images, ideas and words, etc.

Learning is the preliminary step towards the creation of new memory, and memory is related with the information that are retained and recalled in our brain. In learning activity, a potentiation process is utilized, which results in new connections between the neurons, which are utmost important for obtaining and storing new information. Therefore, as new experience or information is obtained, the new connections between the neurons are created within the brain. The repeated recalling or rehearsing results in strengthening of these connections. Humans keep on adapting new and diverse abilities like working on different machines, playing a game, drive a car and walk on the rope, etc. These abilities are obtained through the learning process which is held in the memory. Later, this retained information is used by other brain processes. Table 1 shows different brain parts with their related memory and cognitive tasks. Moreover, brain possesses the ability of adapting novel aptitudes and expertise, retaining the information and reutilizes that information. These retaining and reutilizing abilities regarding aptitudes and expertise are called human memory framework.

Various neuroimaging methodologies like electroencephalography (EEG), positron emission tomography (PET), magneto-encephalography (MEG) and functional magnetic resonance imaging (fMRI) have been used by the scientists in order to investigate these processes. But, EEG and fMRI have many significant features, which are particularly important in neuroscience studies.

Table 1. Brain regions with associated memory and cognitive functions [1, 2].

|    | Brain Parts | Memory Processes |
|----|-------------|------------------|
| 1  | Cerebellum | Restrictive memories, e.g., occasions connected by time |
| 2  | Thalamus | Attention |
| 3  | Frontal lobe | Working and STM |
| 4  | Hippocampus | Initiated to move experiences into memories, e.g., STM |
| 5  | Putamen (temporal, right) | Learning aptitudes, e.g., the capacity to ride a bicycle |
| 6  | Parietal lobe | Spatial memory |
| 7  | Amygdala | Exciting memories |
| 8  | Putamen | Procedural skills |
| 9  | Temporal lobe | Hearing sense |
| 10 | Caudate nucleus | Memories of instinctive skills |
| 11 | Mammillary body | Episodic memory |
| 12 | Central executive (frontal, right) | Holds entire plan, including visual elements |
| 13 | Occipital lobe (visual cortex) | Visual memory |
| 14 | Caudate nucleus (temporal, | Instinctive activities, e.g., grooming is placed here |

Several research studies explored the processes of memory such as encoding, retention and recalling, connected to the responses of brain utilizing brain mapping techniques and memory tests, e.g. fMRI is utilized for brain imaging, and EEG is broadly utilized for brain electrical tests. In encoding, data in short term memory is shifted to long term memory with the help of hippocampus. Basically, hippocampus is the area in human brain incharge of memory arrangement and encoding of information. Hippocampus involvement in encoding of information and its other aspects have been investigated by Battaglia et al. [3]. Retention of information into memory is the ability of human brain to retain data for different time durations, based on the different kinds of memory and content, levels of emotion and attention and redundancies in recall [4]. In case of recall, the information stored within the brain, e.g. content and event, is recalled through reaction to some external stimulus [3].

In this research paper, the objective is to present a review of fMRI and EEG research studies on brain processes which are actively involved in memory retention and recall. Various types of stimuli utilized in cognitive and memory tasks are also discussed. Furthermore, the challenges which are faced in EEG and fMRI methodologies in memory research are also highlighted. The rest of the paper is organized as follows. Section 2 discusses the literature review. Section 3 summarizes our observations and discussion. Section 4 presents the summary and recommendations. Finally, the section 5 concludes the paper.

## 2. LITERATURE REVIEW

In this section various studies, which analyzed EEG and fMRI using ERP component and bands for memory retention and recall, are described. The brain behavior during memory retention and recall by utilizing different tasks are also investigated. Furthermore, impact of presentation modes, content types, stimuli and modalities regarding memory processes are also discussed.

*EEG Studies*: Recently, Event Related Potential (ERP) experiments have been conducted extensively to study short-term memory (STM). Babiloni et al. [5] conducted one such study by analyzing the changes in both the right and left hemispheres of the brain while performing a simple STM task. The subjects were shown a 2D visual-spatial simple sample and were given 4.5 seconds only to memorize the objects. Observations were recorded related with the changes in fronto-parietal theta and alpha rhythms. In another study, the association of event-related potentials (ERP) with the memory and learning processes was investigated by Jongsma et al. [6] through utilizing the multiple trials of digits learning. The developments in memory during the content presentation were observed by the authors. This study utilized sound-related digits as content and then recall task for estimation of memory work, for example, recalling objects in right places in the series. A strong relationship was observed by the authors between ERP component (P300) amplitude and memory work. In the initial STM studies [7], Dunn made use of EEG observations by studying several recognition and memory tasks involving a group of 11 female and 13 male students. The study made use of different levels of cognitive functions, which ranged from recognizing a simple word flashing on a computer screen to more difficult tasks like serial and word category learning. The cerebral areas that were involved in detection of a simple word, single-category memory, and serial-order memory were highlighted on the topomaps of both male and female subjects. A major portion of the research studies regarding memory was in the context of an event related experimental design. The events were utilized as different tasks that showed visual stimuli, and the outcomes were evoked potentials in connection with the tasks / events. In another study, Burgess and Gruzelier [8] carried out research in which they utilized EEG to investigate memory-related changes by performing face and word recognition problems by showing 2D stimuli to the subjects. EEG oscillations corresponding to the alpha and theta bands were analyzed by Klimesch [9]. In this study, the analysis of alpha and theta bands facilitates the study of cognitive and memory functions, wherein an increase in alpha and simultaneous decrease in theta is indicative of a good performance. At the same time, an increase in theta and decrease in alpha band related to the event is also seen in a good performance. Event-related experiments were used by the authors because upper alpha de-synchronization could be seen in these experiments. This upper alpha desynchronization is positively linked to Long-term memory (LTM) performance, on the other hand, theta synchronization has been related to encode new information. The process of consolidating the stored memory takes place during sleep, which is marked by a slow oscillatory activity in the endogenous EEG. Therefore, transcranial slow oscillation stimulation (tSOS; 0.75 Hz) is used to further enhance this process. There has been noted an improvement by the application of this technique. On the other hand, the effect of transcranial slow oscillation stimulation in the awake condition, instead of the sleeping one was studied by Kirov et al [10] and an increase in endogenous EEG slow oscillations (0.4–1.2 Hz) was noted as well. When the subject was in the awake condition, the EEG theta (4–8 Hz) activation was noted to increase due to the transcranial slow oscillation stimulation. This stimulation when applied during learning showed improvements in the memories dependent on the hippocampus area within the brain. However, there was no significant change to the memory consolidation function. The non-linear dynamics of EEG during visual memory functions and its application for research related to memory were studied by Chen et al. [11]. In this study, EEG recordings of 30 physically fit old adults were taken during rest condition, eyes open condition, and picture memory task when 2D multimedia stimulus was shown to them. Calculations were made for the correlation dimension ($D\_2$) for all the subjects. A notable increase was seen in the $D\_2$ while performing memory-related tasks in comparison to the resting state of the subjects. Different regions within the brain were involved as the mental functions varied in complexity. "Interactive 3D Multimedia Learning Tool in Biology – Skeleton" was created by Kamsin & Abdullah [12] as a learning aid for the skeletal system. It allows the user to study the different skeletal system functions interactively. The advantage being that the user has access from all the angles, contrary to the 2D model, in which access is possible only from a single angle. Therefore, as the students can see the skeleton in 3D space, it makes easier to understand and learn. This study made use of a typical 2D display to show the 3D material. In another study, the application of a 3D model to study the process of learning was conducted by Parez and Vanegas [13]. As a result of this study, they concluded that the usage of newer technologies enabled a better communication between the students and teacher, and paves the way for further improvements to the field of education, as compared to the traditional methods of learning and teaching. The 3D material was shown on a conventional 2D display in this study. All the aforementioned studies show that research has been actively done on the factors involved during the learning process which includes memory retention and recall, STM, LTM, and recognition.

Karpicke and Roediger [14, 15] analyzed that retention associated to learn with the testing, which is performed repeatedly. The authors recommended that the process of recalling of retained data from human brain, in which recalling is carried out repeatedly, showed better results of learning and long term rentention of information. During the experiment, subjects were asked to study the list of items under two situations separately. In the first situation, the list of items were observed by the subjects by fifteen times and also tested for five times. While in the second situation, the list of items were looked by the subjects by five times and tested for fifteen times. Then the retention task was performed after the period of seven days and observed that outcome was better for learning and retention for the list of items in the repeated test environment than the repeated study of the list of items. Grady et al [16] utilized words and images as distinctive type of contents and explored the impact of contents on recognition memory in the PFC's right side. The authors found that the quantity of effectively perceived objects larger for imagery contents as compared to words, while the reaction time of the subjects was quicker for images as compared to words. Khairudin et al [17] utilized stimulus of pictures and words for research on the impact of materials on the brain explicit memory. Pictures and words were categorized in negative, positive and unbiased classes for two different analyses. Their outcomes recommended proof of good memory work for pictures than words. The authors also explored the correlation between negative and positive content, which showed that the content of positive nature was recollected in explicit memory better than the negative content. Consequently, the authors found that explicit memory was suppressed by the negative content. In another study, the impacts of visual and verbal content on different processes of memory in episodic memory were explored by Lee et al [18]. They investigated that the tasks related to visual memory were linked with developments in the PFC's right lateral. On the other hand, tasks associated with verbal memory were connected with changes in the blood flow in regional cerebral in PFC's left lateral. The impact of contents of two separate computer games on youngsters was investigated by

Maass et al [19] through analyzing their information & memory performance and learning. The impact of non-violent and violent games were analyzed by author under two separate situations of watching and playing. They concluded that non-violent content had a positive long-term effect on information processing and learning. In another study, six types of content in visual Working Memory like colors, squiggles, polygons, cubes, faces and letters of differing intricacy were analyzed by Eng et al [20]. The result was found that visual Working Memory decreased for complicated visual contents. In another research study, Marois and Ivanoff [21] studied memory limit and processing barriers of the brain on visual STM and underlined three limits: (a) time needed to recognize a visual content, (b) quantity of visual content could be retained in Short Term Memory, and (c) choice of feasible reaction to the content. Diamantopoulou et al [22] utilized distinctive classifications of contents and reviewed memory limitations and retention in this study. They discussed that memory limit, and remembrances were more dependent on likenesses in the content. The behavioral output of the study proposed that expanded reaction times and mistake rates were related with memory load. A deferred facial feeling distinction job was performed by Banko and Vidnyanszky [23] where two confronts, such as, test and sample was utilized like content. The frontal appearances of faces constitute the contents with gradually changing feeling of displaying pleasure. Presentation of two faces was made for a duration of 300 ms with two separate periods of retention for one or six seconds. The process of STM for facial attributes was found by the authors and observed that the period of storing the information alters the processes of the human neural system in the procedure of STM encoding. The impact of moderate and quick learning methodologies was examined on retention of information by Joiner & Smith [24] and observed the slow process of learning added firmly to LTM as compared to quick learning methodology. In [25], author examined the cortical responses on human memory by utilizing a 2D easy visuo-spatial test of one bit to be remembered in case of STM. Two vertical bars were utilized as a content in the test. It was noted that time taken by the subjects for a response was delayed from 3.5 to 5.5 seconds. Authors also examined that there was no statistical variation among the two conditions. However, EEG outcomes showed that theta band (4 to 6 Hertz) curtailed in left frontal and bilateral parietal areas during the delayed time. Moreover, it was also found that high alpha band (10 to 12 Hertz) was declined in the left regions of fronto parietal and low alpha band (6 to 8 Hertz) was reduced in left parietal and bilateral frontal parts. The author observed that a simple STM test produced the variations in the alpha and theta band at the fronto-parietal part. Based on single-trial EEG activity, Eunho Noh et al. [26] presents a research study to investigate the performance of subsequent memory during and before appearance of content for the subject. The main idea is to carry out two-class based classification for prediction of subsequently forgotten vs remembered trials in the light of subject's reactions in the recognition stage. Their study gives an illustration from the analysis to recognize the qualities of the different SMEs before and during content presentation based on single trial.

*fMRI Studies*: A fMRI study was carried out by Karlsgodt et al [27] through utilizing verbal WM task. During the process of encoding, retention and recall of information from human brain, subject examination proposed that like LTM, the hippocampus played a similar role in WM. This suggested that hippocampus may also be included into the process of encoding and recalling instead of only LTM tasks. In another study, Crottaz-Herbette et al [28] investigated that modality impacts on verbal WM. They utilized fMRI technique with closely resembling content and explored the likenesses and fluctuations between visual verbal WM and sound-related verbal WM. They analyzed similarities and contrast during visual and sound-related tasks in the cerebrum. The examination showed huge modality impacts in the initiation of the parietal and prefrontal cortex. Moreover, higher activity was observed in sound-related verbal WM indicated by left dorsolateral prefrontal cortex (PFC). On the other side, the left posterior parietal cortex showed greater activation in visual verbal WM than the sound-related verbal WM. Right hemisphere did not contain any such changes. Various studies of fMRI have revealed a strong association between the brain mechanisms and its initiation in the regions of pari-etal, medial temporal, occipital and prefrontal cortex (PFC). The authors found that prefrontal cortex was actively involved to have unique activation appearances for different cognitive tasks like verbal and spatial working memory, semantic memory, word identification, episodic memory, constant attention, understanding and smell [29]. In another study [30], authors examined the retrieval of information from LTM in an investigation that comprised of two stages, encoding and retreival, utilizing the fMRI. The authors proposed that to retrieve the information from the human memory, left dorsolateral prefrontal cortex may be considered. In another study [31], authors investigated the performance of WM by utilizing three dissimilar contents for visual WM tasks like shape and colour, only colour and only shape. They observed that occipito-temporal, frontal and parietal regions within the brain are actively involved in the tasks. They found that prefrontal region was responsive to load handling only, and posterior parietal cortex was responsive to memory load handling as well as to feature's identification with respect to their complexity like shapes. In another fMRI study [32], authors examined the information encoding state in visual WM by utilizing two separate classes of contents i.e. faces and colour. They observed a large dissimilarity in encoding time taken by the subjects. They found that in case of face contents, much information was stored into visual WM. HHHowever, in case of colour contents, reaction time was much quicker. Michels et al. [33] used Sternberg WM test to analyze the low and high EEG frequencies. They observed that theta band (5 to 7 Hz) is increased significantly in the frontal part of the brain. On the other side, beta 1 (13 to 20 Hz), a few regions in the beta 2 (20 to 30 Hz), gamma and alpha band 1/2 indicated high frequencies in the parieto-occipital parts. They also found during the analysis based on the load in the retention phase that all bands excluding alpha 1 demonstrated positive increment with set size. Authors also investigated the linkage between the fMRI Bold Signal and EEG bands in the Working Memoery test during the retention phase and explored that low and high frequencies are associated with fMRI Bold signal.

## 3. OBSERVATIONS AND DISCUSSION

From the above literature review, we discuss our observations in this section which are related to stimulus effect on different memory processes, elements effect on memory retention and recall, identification of brain regions which are activated in EEG experiments during cognitive tasks w.r.t alpha and theta band, brain regions activated in fMRI experiments during cognitive tasks and challenges related to fMRI and EEG Brain Mapping Techniques. Table 2 presents our findings which are observed by studying stimulus effect on different memory processes.

**Table 2. Stimulus effect on different memory processes. [WM (Working Memory), STM (Short-Term Memory), LTM (Long-Term Memory)].**

|   | Study | Stimulus | No. of Samples | Avg. Age (Yrs) | Memory Processes | Observations / Findings |
|---|-------|----------|----------------|----------------|------------------|-------------------------|
| 1 | [34] | Negative and neutral stimuli i.e. pictures like animals and people | 46 | 23.7 | LTM and WM | Significant finding which is observed is that LTM retains negative information more profoundly than neutral information. Another important observation is that the WM is not consistently affected by the emotional content but only in few cases. |
| 2 | [35] | Visual stimuli like neutral faces, happy faces and squares | 20 | 25.5 | Visuospatial working memory | It is observed that trials in which happy faces are used, result in greater retention of information as well as the correct responses. |
| 3 | [36] | Real objects and line drawings | 96 | | Recognition memory | The results give rise to a key observation that while making discrimination between left and right ATL groups, the results for the objects were better than line drawings. However, as far as the classification results are concerned for the candidates, the objects were better classified than line drawings for the left ATL groups, on the contrary, the classification results for objects and line drawing are same for right ATL groups. |
| 4 | [37] | Neutral, emotional and negative words | 18 | 21.5 | Memory and learning | It is observed that data processing as well as learning is positively affected by emotional contents as compared to the negative content which has a negative impact on learning as well as data processing. |
| 5 | [38] | Geometric items | 12 | 26.1 | Encoding and Recall | Both segregation and lateralization based on the explicit relational features of stimuli requiring the intra-item connections for retrieval and encoding. |
| 6 | [39] | Scene images (indoor & outdoor) | 15 | | Encoding and Recognition memory | MTL sub regions are observed to have content specific source encoding patterns of activation. These give rise to the conclusion that stimulus category has the greater effect on MTL's memory effects. |
| 7 | [40] | Faces, Scenes, objects | 34 | 23.2 | Memory and learning | Increased perirhinal activity is observed when the subjects distinguished among objects and faces showed from different viewpoints instead of identical views. On the other side, it is also observed that posterior hippocampus presented an effect of viewpoint for both scenes and faces. These observations give evidence about the involvement of medial temporal lobe (MTL) in processes other than long-term declarative memory. |
| 8 | [41] | Violent Vs non-violent PC game | 167 | 17.6 | Memory and learning | Emotional content has a positive long-term impact on data processing and learning as compared to negative content, which has a negative impact on learning and data processing. |
| 9 | [42] | Images and words | 12 | 25.6 | Recognition memory | Images are identified more than words with quick response |
| 10 | [43] | Images and words | 15 | 18 to 32 | Recognition memory | Depending upon the nature of the content's material, activities associated with source memory varied. |

Table 3 shows our observations related to elements which effect on memory retention and recall operation as mentioned below:

**Table 3. Elements' effect on memory retention and recall operation.**

| | Study | Effect on Human Memory | Observations and Discussion |
|---|---|---|---|
| | | **Reward** | |

| # | Study | | |
|---|---|---|---|
| 1 | J.P. O'Doherty et al [44] | Non-reward learning or discipline circumstances do not produce good results as compared to reward-based learning. | During a certain assignment, rewards may attract more attention and keep the neuron associations with each other. Due to this effect, it produces constant performance on the memory. |
| **Effect of Testing** | | | |
| 2 | Andrew C. Butler [45] | Testing reduces the error rate in recalling outcome and also holds the correct data. Additionally, it allows to recall the retained information from the human memory. | Frequent testing activates the same neurons in the begining, to learn how to execute more than once. Due to the testing effect, neurons are more harmonized and steady. It helps to gather the retained data from the brain in a simple way. |
| **Mnemonics** | | | |
| 3 | Todd M. Franke and Joel R. Levin et al [46] | Mnemonics help encoding to memorize difficult data in easy way and recalls such data accurately. | It makes convenient to store and retrieve complex information from the brain and also make stronger network connections. |
| **Nutrition and Exercise** | | | |
| 4 | Kasper Skriver et al [47] | Nourishment and physical activity produces a lot of advantages for mental health and have a larger impact on neuro-transmitters. | Consistent physical activity and balanced food keep the cerebrum of old persons from contracting and also assist to retain the memory in good condition. Regular physical activity and balanced nutrition have important impact on the neurotrophic components and plasticity in cerebrum areas related to memory performance. |
| **Sleep** | | | |
| 5 | Carmen E. Westerberg et al [48] | Sleep improves memory association and cerebrum plasticity. | It is still undetermined that how sleep generates the changes in the nervous system which influences the consolidation procedure in the human brain. |
| **Attention** | | | |
| 6 | D. Li et al [49] | Limited attention minimizes the recalling performance. On the other side, full attention maximizes the memory recall and retention of information in the brain. | The ability of retaining the information in the working memory is reduced due to divided attention and less concentration. Due to this, WM cannot retain the information which is needed. Therefore, the outcome is incomplete memory union, which finally decreases the recalling performance. |
| **Rehearsal** | | | |
| 7 | Svoboda and Brian Levine et al [50] | Recall performance is improved by regular rehearsal. | Snaptic associations between the neurons, involved in a certain assignment are strengthened due to regular practice. Resultantly, the stronger synaptic connections between the neurons help in improved recollection of subsequent information from the memory. |

Several research studies show that encoding, retention and recall activities are mostly related to alpha and theta bands by making variations in EEG experiments during the STM, WM and LTM tasks. Table 4 presents observations related to identification of the brain regions which are activated in EEG experiments during cognitive tasks.

**Table 4. Identify brain regions which are activated in EEG experiments during cognitive tasks w.r.t Alpha and Theta Band.**

| Study | Variations in EEG | Brain Area | Category |
|---|---|---|---|
| **Alpha Band** | | | |
| **Encoding** | | | |

| | | | | |
|---|---|---|---|---|
| 1 | Uwe Friese et al [51] | Higher Alpha activity | Right frontal area | WM |
| **Retention** | | | | |
| 2 | M.D. Rugg et al [52] | Reduced Alpha activity | At temporal, central and left hemisphere parietal regions. At occipito-parietal and frontal areas. | STM |
| 3 | Ole Jensen et al [53] | Increased Alpha power | Midfrontal (Cz, Fz) parietal–occipital cortex | WM |
| **Theta Band** | | | | |
| **Encoding** | | | | |
| 1 | Daria Osipova et al [54] | Increases Theta [5 to 10 Hz] power | At Parietal-occipital area (Pz, POz, O1 and O2,). At Frontal area (AFz, Fz, Fp1,Fp2,AF3,AF4,F3, F4, F7 and F8) | STM |
| 2 | Uwe Friese et al [51] | Increased Theta activity | Right frontal area | WM |
| **Retention** | | | | |
| 3 | M.D. Rugg et al [52] | Increased Theta activity | At temporal, central and left hemisphere parietal regions. At occipito-parietal and frontal areas. | STM |
| 4 | Ole Jensen et al [53] | Increased Theta power | Midfrontal (Cz, Fz) parietal–occipital cortex | WM |
| **Recall** | | | | |
| 5 | Per B. Sederberg et al [55] | Increased Theta power and modulated with task complexity | Frontal area | WM |
| 6 | Marcel C.M. Bastiaansen et al [56] | Increased Theta [4 to 7 Hz] power | Left occipital area | WM |

Table 5 shows our obervations related to identification of brain regions which are activated during the cognitive tasks by performing various fMRI experiments:

**Table 5. Brain regions activated in fMRI experiments during cognitive tasks.**

| | Study | Brain Area | Memory Type |
|---|---|---|---|
| **Encoding** | | | |
| 1 | Heiko C. Bergmann et al [57] | Medial Temporal Lobe | LTM |
| 2 | Max Toepper et al [58] | Hippocampal | WM |
| **Retention** | | | |
| 3 | J. Jay Todd et al [59] | posterior parietal cortex | WM |
| 4 | Donohue et al [60] | Posterior middle temporal gyrus and Ventrolateral prefrontal cortex | LTM |
| **Recall** | | | |
| 5 | Todd S. Braver et al [61] | Frontopolar Prefrontal Cortex | LTM |

| | | | | |
|---|---|---|---|---|
| 6 | Manenti et al [30] | | Dorsolateral prefrontal cortex | LTM |

Table 6 highlights the issues and challenges with EEG and fMRI procedures in memory research.

**Table 6. Problems and Difficulties related to fMRI and EEG Brain Mapping Techniques.**

| | Research Study | Technique | Issues | Challenges |
|---|---|---|---|---|
| 1 | S. Haller et al [62] | fMRI | Design | Head motion is a critical limitation in the experimentation as it can cause in variations in the intensity of changes in fMRI signals. Another challenge is that if the maximum activation is very close to the baseline property, then some processes will not have meaningful representation. |
| 2 | G. Aguirre et al [63] | fMRI | Block related design | The challenges faced include the head motion which occurs especially in the case where only few blocks are used. Baseline condition problem is also a major issue as its poor choice gives rise to meaningless conclusions. Some of the other issues include the repetition of several tasks because in each block, one condition is used only so the stimulus cannot be randomized in a block which in turn makes the stimulus predictable, and the person in the experiments get to know the events. |
| 3 | BR Rosen et al [64] | fMRI | Event related design | The major problem is the statistical design of the experiments as the variation in fMRI signal as single signal presentation is not enough. |
| 4 | Stark and Squire et al [65] | fMRI | Baseline versus activity conditions | The baseline conditions play a critical role in fMRI and the ambiguous baseline conditions pose a bigger problem during rest in fMRI as compared to other alternate baseline conditions. The aim of this activity in rest is to produce calmness in order to reverse the effects of activity, which occurred during tasks related to memory functions. However, after analysis, it becomes evident that during rest the cognitive activity still occurs significantly, therefore, it is unsuitable for baseline. For the better understanding of baseline and activation conditions, a lot of data needs to be examined and interpreted, e.g., breathing, etc. |
| 5 | Poldrack RA et al [66] | fMRI | Reverse inference | fMRI helps in the measurement of activation of various parts of brain, which are activated due to the cognitive tasks. The resulting data of this process helps neuroscientists to get valuable information about the role of various parts of brain in cognitive function. A problem often faced is about the identification of brain regions during the activation of labelled cognitive processes. To resolve this issue, the reverse inference is used, which backtracks from the occurrence of brain activation to some particular cognitive function. |
| 6 | Poldrack RA et al [66] | fMRI | Forward inference | The challenge faced in forward inference is due to its correlation nature, which leads to uncertainty about the activation of brain regions in the cognitive process, i.e., whether those are important for the execution of processes or not. |

| 7 | Roberta Grech et al [67] | EEG | Inverse problem in source analysis | The major challenges which are difficult to meet include determining the areas within the brain which have a critical role with respect to EEG waves of interest and which are responsible for the recording of scalp potential. |
|---|---|---|---|---|
| 8 | Hans Hallez et al [68] | EEG | Forward problem in source analysis | The challenges of the forward problem are met by calculating the potentials at the electrodes in response to given electrical sources. |
| 9 | Kondylis et al [69] | EEG | poor spatial resolution | Post synaptic potentials which are generated by superficial layers of the cortex have the capability to influence EEG signals. On the contrary, the dendrites which are in the deep areas of cortex have not much influence on the EEG Signals. |
| 10 | Hamalainen et al [70] | EEG | Do not directly capture axonal action potentials | Current quadrupole efficiently represents an active potential which gives rise to the drastic decrease in the resulting field as compared to those which are created by the current dipole of post synaptic potentials. As EEGs are resultant of average of large quantity of neurons, therefore, it is evident that a synchronous activity involving large quantity of cells must be in order to have significant effect on the recordings. |

## 4. SUMMARY AND RECOMMENDATIONS

Various studies of memory types and processes are reviewed in this research paper. Our main focus is on two vital human memory processes i.e., memory retention and recall. The conventional models investigated are long and short-term memory, phenomenon of working memory, visual and audio data preparation, better utilization of cognitive capacity and cerebrum channel's data processing of auditory and visual information. ERP component (P300) is closely related to memory performance like recalling of information from the memory. Retrieval of information is associated with the right parietal cortex and encoding processes are associated with left parietal cortex due to the prominent gamma band. Simple STM task brings changes in the alpha and theta frequencies at the frontal-parietal area. Moreover, output is better for learning and retention for the list of items in the repeated test environment than the repeated study of the list of items. Also, hippocampus is playing an important role in overall information encoding and recalling process in contrast to simply in LTM tasks. Left dorsolateral prefrontal cortex may be considered for retrieval of information from the human memory. Memory recall is successfully differentiated at left temporal lobe through EEG gamma band and parietal and frontal brain regions through EEG theta band. Activated temporal and frontal brain regions show that multimedia content plays an important role in memory retention and recall, as brain receives both visual and verbal information during learning phase. In this case, brain creates a new memory by combining both the modalities i.e., visual and audio. EEG low frequency band at parietal and frontal regions and high frequency at temporal region show successful memory recall. Several regions of the brain, particularly amygdala, cerebral cortex and hippocampus are firmly connected with memory processes like retention and recall. The correct memory information retrieval, working and episodic memory are significantly influenced by the cerebral cortex in the temporal, frontal and parietal lobes. Hippocampus is the area in the medial temporal lobe responsible for movement of data from short-term to long-term capacity, development of new memory and memory encoding. Various human emotions are linked with amygdala activation like happiness, fear and sadness, etc. New memory formation, and learning is strongly affected by the long-term potentiation. The strong and weak potency of the human memory is based on the synaptic communication network power. The process of correct recall and retention of new data into the memory is highly associated with strengthen synaptic con-nections. Research has been carried out to identify the factors which have a critical role to retain much information and recall from the memory after a longer duration of time. One of these factors like sleep, attention and rehearsal, etc., are proved analytically. These factors play their important role in using the restricted WM space and learn further information and shift to LTM capacity. However, these all factors may not be involved in high recall or retention processes. However, each factor plays an important role in specific situation.

In the existing research work carried out by the research community, the work has not been done upon the effect of 2D / 3D educational content on memory retention and recall. Therefore, the research can be carried out to study the brain behavior during learning and memorization process for evaluation on the effects of 2D / 3D educational contents on the memory retention and recall. Also, brain regions can be identified by its various generated brain signals. For this purpose, backtracking techniques can be used, for example, particle swarm optimization (PSO) algorithm, backtracking search optimization algorithm (BSA) or to solve the real-valued numerical optimization problems, a new evolutionary algorithm (EA). In order to solve the above mentioned complex numerical optimization, non-differentiable and non-linear problem, backtracking search optimization algorithm (BSA), which is a famous stochastic search algorithm, can be used. Also from the research studies [53, 71-74], alpha band can be considered as an important analysis component for information retention and recall from human brain. Therefore, it can be used to identify the brain regions which are activated during cognitive tasks in EEG experiments. Research can also be carried out for classifying the learning process as positive (good memory retention) or negative (bad memory retention). Furthermore, the learning process can be analyzed from the features obtained from EEG scalp maps. Also more invasive and sophisticated recording techniques may be used for recording the signals from the scalp in order to avoid the noise and artifacts.

## 5. CONCLUSIONS

In this paper, we review the best-known research studies of EEG and fMRI related to memory processes, which are involved in memory retention as well as recall along with the various types of contents used in cognitive and memory tasks. EEG is a detailed analysis tool which is used to record the electrical activities on the scalp produced by brain. It is very useful due to its high temporal resolution and direct measurement of neurons' activity of the brain. Furthermore, EEG high and low frequency domains have been noted for research purposes by utilizing diverse nature of stimuli for retention and recall processes in STM, WM and LTM tasks. Also, fMRI becomes very useful to localize the activity of the brain precisely in millimeters. The memory paradigms, which are being used by scientists for the stimulation of various brain regions during cognitive tasks, are also discussed to understand the deep brain and cortical cognitive abilities. Posterior parietal cortices, bilateral hippocampal, hippocampus, prefrontal, fronto-central, frontal and occipital-parietal regions are activated for memory retention and recall processes during the memory cognitive tasks. The brief overview of experimental design factors will give a useful direction to the researchers to design intelligent cognitive tasks in future research. Furthermore, the key problems and challenges being faced by memory research of brain mapping methodologies, i.e., EEG and fMRI, are also discussed in this paper. Recommendations for memory research have also been suggested. We are hopeful that this review will provide worthwhile understandings and insights for memory researchers.

**Compliance with ethical standards**

**Conflicts of interest:** The authors have no conflicts of interest to declare.

**Ethical approval:** All procedures performed in studies involving human participants were in accordance with the ethical standards of the institutional and/or national research committee and with the 1964 Helsinki declaration and its later amendments or comparable ethical standards.